\documentclass[reprint,groupedaddress]{revtex4-2}

\usepackage{graphicx}
\usepackage{upgreek}
\usepackage{amsmath, amssymb}
\usepackage{lipsum}
\usepackage{stackengine}
\usepackage[normalem]{ulem}

\usepackage{color} % for the command \textcolor
\usepackage{soul} % for the command \hl
\usepackage{hyperref}
\hypersetup{
    colorlinks=true,
    linkcolor=blue,
    filecolor=magenta,      
    citecolor=blue,      
    urlcolor=blue,
    }
\DeclareMathAlphabet\mathbfcal{OMS}{cmsy}{b}{n}

\newcommand{\ro}{\mathbf{r}_0}

\newcommand{\cj}{^{*}}

\newcommand{\sw}{(\mathbf{r}_{0},\omega)}
\newcommand{\avg}[1]{\left\langle #1 \right\rangle}

\newcommand{\im}[1]{\text{Im}\left[#1\right]}
\newcommand{\re}[1]{\text{Re}\left[#1\right]}

\newcommand{\nx}{\hat{\mathbf{x}}}
\newcommand{\ny}{\hat{\mathbf{y}}}
\newcommand{\nz}{\hat{\mathbf{z}}}
\newcommand{\nr}{\hat{\mbf{r}}}

\newcommand{\E}{\mathbf{E}}

\newcommand{\Eo}{\mathbf{E}_0}

\newcommand{\Es}{\mathbf{E}_s}

\newcommand{\dE}{\delta\mathbf{E}}

\newcommand{\wo}{\omega_0}

\newcommand{\II}{\overline{\overline{\mathbf{I}}}}
\newcommand{\G}{\overline{\overline{\mathbf{G}}}}
\newcommand{\Go}{\overline{\overline{\mathbf{G}}}_0}
\newcommand{\Gs}{\overline{\overline{\mathbf{G}}}_s}

\newcommand{\dyad}[1]{\overline{\overline{\mathbf{#1}}}}

\newcommand{\ft}[1]{\tilde{#1}}
\newcommand{\atrzero}{\bigg|_{\mbf{r}_{0}=\mbf{r}'_{0}}}

\newcommand{\mbf}[1]{\mathbf{#1}}

\newcommand{\conv}[2]{\left( #1 \ast #2 \right)(\mathbf{r}_{0},\omega) }

\renewcommand{\d}[1]{\ensuremath{\operatorname{d}\!{#1}}}

\begin{document}

\title{Backaction suppression in levitated optomechanics using reflective boundaries}

\author{Rafa{\l} Gajewski}
\author{James Bateman}
\email[]{j.e.bateman@swansea.ac.uk}
\affiliation{Department of Physics, Faculty of Science and Engineering, Swansea University, SA2 8PP UK}

\date{\today}

\begin{abstract}
  We show theoretically that the noise due to laser induced backaction acting on a small nanosphere levitated in a standing-wave trap can be considerably reduced by utilising a suitable reflective boundary. We examine the spherical mirror geometry as a case study of this backaction suppression effect, discussing the theoretical and experimental constraints. We study the effects of laser recoil directly, by analysing optical force fluctuations acting on a dipolar particle trapped at the centre of a spherical mirror. We also compute the corresponding measurement imprecision in an interferometric, shot-noise-limited position measurement, using the formalism of Fisher information flow. Our results show that the standing-wave trapping field is necessary for backaction suppression in three dimensions, and they satisfy the Heisenberg limit of detection.
\end{abstract}

%\keywords{}

\maketitle
\section{Introduction}
In recent years, experiments which utilise nano-sized particles trapped at the focus of a laser beam emerged as a promising experimental platform for detecting ultrasmall forces \cite{zeptonewton_2016,Ahn_2020}, gravitational waves \cite{aggarwal_2022}, searching for dark matter \cite{monteiro_2020}, and potentially probing quantum behaviours of large masses in wide and untested parameter regimes \cite{Kaltenbaek_2023}. These levitated optomechanical systems constitute high quality mechanical oscillators, whose position is probed by phase-sensitive measurement of the light scattered by the particle. The interferometric position readout and extreme isolation from thermal dissipation \cite{levitodynamics_2021}, reached due to the lack of physical contact, enabled levitated optomechanical systems to cool the particle motion to the ground state. \cite{Delic_2020,Magrini_2021,Piotrowski_2023}, with recent experiments showing prospects for quantum-enhanced sensing below the shot-noise level \cite{Militrau_2022}.

However, in a continuous measurement by the laser, fluctuations of the electromagnetic field introduce random perturbations to the motion of the mass known as measurement backaction \cite{Braginsky_1977,caves_1980, clerk_2010, Jain_2016}. This backaction noise is linked to the minimum achievable imprecision in the position measurement by the Heisenberg limit of detection \cite{clerk_2010,tebben_2019}, which keeps the product of the two quantities fixed. Although cooling of the particle to its motional ground state has been demonstrated \cite{Delic_2020,Magrini_2021,Piotrowski_2023}, the coherence times of the prepared motional states are limited by excessive recoil heating imposed by the optical measurement process. This limits access to the envisioned applications \cite{levitodynamics_2021} and therefore finding ways to mitigate the effects of backaction is of vital importance.

The degree to which the backaction process can be controlled in levitated optomechanics without turning off the trapping light has not yet been fully explored. Recently, a backcation suppression method has been proposed based on illumination by squeezed light \cite{ballestero_2023}, enabling control over the information encoded in the radiation scattered by the particle. A different method relies on trapping hexagonal plates instead of spheres \cite{aggarwal_2022,winstone_2022}, because a different geometry of the trapped object alters its radiation pattern, thereby confining the solid angle in which backaction noise can act.

Backaction noise in levitated optomechanics can be understood as a result of interference of the laser field with the local vacuum field fluctuations \cite{caves_1980,tebben_2022,sinha_2020,sideband_2022}.  In general, the vacuum fluctuations are function of the environment, depending on its material properties and geometry \cite{argwalI,novotny_2012principles}. It is known that surfaces can lead to backaction different than would be expected in free-space \cite{sinha_2020}. In this work, we investigate an experimental scheme to suppress backaction by trapping the particle at the centre of a spherical mirror. We show that a spherical geometry in conjunction with a standing wave optical trapping potential can lead to a substantial reduction in mechanical noise, limited mainly by the mirror reflectivity. 

Previous QED analysis of the spherical mirror geometry found that the local density of optical states is strongly modify at the centre \cite{Hetet_2010} which allows for full suppression of spontaneous emission from an atom. In the context of levitated optomechanics, optimal position measurement of a particle trapped at the centre of a spherical mirror was recently analysed \cite{selfhomo_2021,selfhomo_2022} and the authors show that this can be used to realise an ideal reference field for self-homodyne detection which is theoretically able to reach the Heisenberg limit of detection by observation from only half of the solid angle. 

A similar setup was also recently proposed for the purpose of backaction suppression \cite{weiser_2025}. While our findings do not contradict the general protocol presented in this recent paper, we find that suppression of scattered power \textit{does not} correspond to the suppression of backaction. Instead, we find that the linear position read-out becomes inaccessible in an interferometric measurement of the far-field scatter when the scattered power is \textit{maximally enhanced} by the spherical mirror. Our results suggest that the scheme of constraining backaction noise presented in \cite{weiser_2025} could be successfully applied with a small adjustment of the spherical mirror radius. We also find that full backaction suppression effect in three dimensions is possible when the spherical mirror covers half of the solid angle and the light probing the position of the particle corresponds to a maximum of an intensity standing wave. We show that our results satisfy the Heisenberg limit of detection, and therefore constitute a trade-off between measurement imprecision and backaction without altering the illuminating laser power. This work has general implications for recoil noise reduction in a structured environment, offers a feasible experimental implementation and may aid in the search for other useful configurations.

\section{Problem statement}
\begin{figure}
    \centering
    \includegraphics[width=\linewidth]{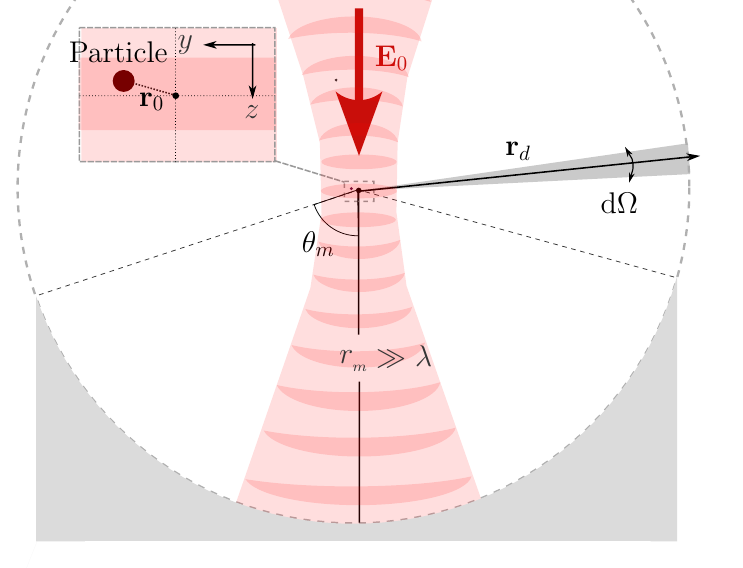}
    \caption{Spherical mirror trapping configuration. A laser beam polarised out of the page is focused close to the centre of the spherical mirror, such that the beam retroreflects from the mirror surface and generates a standing wave trap. A particle is trapped at the central maximum with displacement from the centre $|\mbf{r}_0|\ll\lambda$.}
    \label{fig:swtrap}
\end{figure}

We consider the setup shown in figure \ref{fig:swtrap} as a case study.
In this setup, the beam is incident upon the mirror surface and forms a standing wave trap. We assume that the beam parameters are selected such that the beam is retroreflected and an intensity maximum appears at the mirror centre (see appendix \ref{sec:retro}) which we also take to be the origin of the coordinate system.
The particle is trapped at the intensity maximum appearing at the centre of a spherical mirror, and is located at $\mbf{r}_{0}$ (with $|\mbf{r}_0|\ll\lambda$) which corresponds to the displacement due to the motion about the trap centre. The particle is modelled as a point electric dipole \cite{novotny_2012principles} (radius $r_p\ll \lambda$) with isotropic polarisability $\alpha$ acted on by a stochastic force $\delta\mbf{F}$. We assume that we can ignore the contribution of residual gas and that the force fluctuations are dominated by the contribution from laser recoil. Throughout this analysis, we work in the quasistatic limit, ignoring any effects the motion has on the imprecision-backaction characteristics. Assuming that the mirror is not too large, this is permissible since the particle motion is significantly slower than time delay of the mirror-particle round trip.

We assume that the beam is weakly-focused and that the central maximum is sufficiently close to the beam waist such that the standing wave field resembles a field of two counter-propagating plane waves,
\begin{equation}
    \Eo(\mbf{r},t)=\re{E_{0}\cos(Akz)e^{-i\omega_0 t}}\nx.
\label{SWz}
\end{equation}
In equation \eqref{SWz}, $k=\omega_0/c=2\pi/\lambda$ is the free-space wavenumber of the laser beam and the Gouy factor $A$ increases the effective wavelength of the field, depending on the degree of focusing \cite{tebben_2019}. We note that the standing wave as shown in figure \ref{fig:swtrap} depends on the mirror radius. We address the question of maintaining an intensity maximum (and therefore the equilibrium trapping position) at the centre of the mirror in appendix \ref{sec:retro}.

In the next section, we compute measurement backaction by employing a stochastic electrodynamics description of the local field fluctuations, \cite{sideband_2022, tebben_2022, friction_2004,Henkel_2002,agrenius_2023} and computing the spectral noise density of the recoil force  in a standing wave with an intensity maximum at the mirror centre, \textit{independent of how the standing wave was formed}. The results of section \ref{sec:backaction} and appendix \ref{sec:retro} show that the condition on the mirror radius required to keep an intensity maximum of a standing wave at the centre coincides with the condition corresponding to full suppression of backaction noise. Therefore, the setup shown in figure \ref{fig:swtrap} shows how our results can be applied in an experiment to demonstrate suppression of backaction in three dimensions.

\section{Measurement backaction}\label{sec:backaction}
In this section we quantify the measurement backaction by computing the spectral density of the optical force fluctuations at the mechanical frequency of the trapped particle. In previous studies, recoil was inferred from the radiation pressure that the dipole emission would exert in the far-field \cite{tebben_2019,selfhomo_2021,weiser_2025}. However, when reflective boundaries are present, it is unclear whether this approach is valid, since the direction of recoil depends on how the scattered radiation maps to its reflection in the far-field. Here, we instead compute the optical force directly at the location of the particle.

The particle interacts with the laser field and a fluctuating background field,
\begin{equation}
	\E(\mathbf{r}_{0},t) = \Eo(\mathbf{r}_{0},t) + \dE(\mathbf{r}_{0},t).
\label{total field stoch}
\end{equation}
In equation \eqref{total field stoch}, $\dE$ represent zero-point or thermal field fluctuations that are modelled as a statistically stationary process with zero average \cite{sideband_2022}, and $\Eo$ is the trapping laser field. In general, the particle also interacts with reflections of its own scatter, which lead to an infinite series of additional terms in equation \eqref{total field stoch} scaling in powers of $k^3\alpha$. Since $k^3\alpha\sim(r_{p}/\lambda)^{3}\ll 1$ this series converges quickly, and since we are interested in computing the dominating contribution to backaction, we truncate at zeroth order in $\alpha$. Treating the particle as a point electric dipole located at $\mbf{r}_{0}$, the field in equation \eqref{total field stoch} induces a dipole moment $\mbf{p}$. In this dipole approximation, the particle experiences an optical force given by, \cite{novotny_2012principles}
\begin{equation}
	\mbf{F} = \sum_{i} p_i\nabla E_i + \frac{d}{d t}(\mbf{p}\times\mbf{B})
	\label{optical force}
\end{equation}
where $i=x,y,z$. Since we are only interested in computing the contribution of laser recoil, we ignore the intrinsic fluctuations of the dipole moment. In the frequency domain in the limit $\omega\ll\omega_0$, the force is well approximated by the first term in Eq.~\eqref{optical force}, 
\begin{align}
	\ft{\mbf{F}}\sw &= \sum_i\conv{\ft{p_i}}{\nabla \ft{E_i}}\nonumber\\
									&\approx \tilde{\mbf{F}}_{0}\sw + \delta \tilde{\mbf{F}}\sw
	\label{force simple}
\end{align}
where the quantities with a tilde denote the respective Fourier transforms and the `$\ast$' symbol denotes a convolution in frequency.
In the second line of equation  \eqref{force simple} we chose $\tilde{\mbf{F}}$ to represent the deterministic force and $\delta\tilde{\mbf{F}}$ to represent force fluctuations, approximated with terms containing fluctuations in the fields up to linear order,
\begin{equation}
	\delta \ft{\mbf{F}}\sw = \conv{\alpha\tilde{E_{0}}}{\nabla\delta\tilde{E}} + \conv{\alpha\delta\tilde{E}}{\nabla\tilde{E_{0}}}  ,
\label{force fluctuations general}
\end{equation}
where we used $\ft{\mbf{p}}=\alpha\ft{\mbf{E}}$ and have assumed that $\mbf{E}_{0}$ is $\nx$ polarised, which allows us to drop the subscript `$x$' from the field components $\delta E_{x}$ and $E_{0x}$ for clarity. Assuming that $\delta\ft{\mbf{F}}$ is a stationary stochastic process with zero ensemble average, we find the corresponding spectral density of the force fluctuations using the expression, \cite{mandel_wolf_1995}
\begin{equation}
	\label{sd indices}
	S^{F}_{ii}(\omega)=\int_\mathbb{R}\avg{\delta \tilde{F}_i\cj\sw \delta \tilde{F}_i(\mbf{r}_{0}',\omega')}\d\omega'\atrzero,
\end{equation}
where the angled brackets denote an ensemble average. We can evaluate the field correlations arising in equation \eqref{sd indices} by applying the fluctuation-dissipation theorem for the field fluctuations given in equation \eqref{fdt}. Upon substitution of equation \eqref{force fluctuations general} and \eqref{fdt} into equation \eqref{sd indices}, we find that for a monochromatic driving field of the form $\mbf{E}(\mbf{r},t)=\re{\ft{E}(\mbf{r})e^{-i\omega_0t}}\hat{\mbf{x}}$, the spectral density of force fluctuations in the limit $\hbar\omega_0\gg k_B T$ can be approximated as, \footnote{See the derivation in section 2 of the supplementary material for more details, where we also recover the free-space backaction expressions.}
\begin{align}
	S^{F}_{ii}(\omega) &= \frac{\hbar\omega_{0}^{2}}{4\pi c^{2}\epsilon_0}\alpha^{2}\Big(\nonumber\\
										 &\phantom{+}\ft{E}_{0}\cj(\mbf{r}_{0})\ft{E}_{0}(\mbf{r}_{0}')\partial_i\partial_i'\im{G_{xx}(\mbf{r}_{0},\mbf{r}_{0}',\omega_{0})}\nonumber\\
										 &+\partial_i\ft{E}_{0}\cj(\mbf{r}_{0})\partial_i'\ft{E}_{0}(\mbf{r}_{0}')\im{G_{xx}(\mbf{r}_{0},\mbf{r}_{0}',\omega_{0})}\nonumber\\
										 &+\ft{E}_{0}\cj(\mbf{r}_{0})\partial_i'\ft{E}_{0}(\mbf{r}_{0}')\partial_i\im{G_{xx}(\mbf{r}_{0},\mbf{r}_{0}',\omega_{0})}\nonumber\\
										 &+\ft{E}_{0}(\mbf{r}_{0}')\partial_i\ft{E}_{0}\cj(\mbf{r}_{0})\partial_i'\im{G_{xx}(\mbf{r}_{0},\mbf{r}_{0}',\omega_{0})}\Big)_{\mbf{r}_{0}=\mbf{r}_{0}'}
\label{sd mono final}
\end{align}
\begin{figure}
	\centering
		\includegraphics[width=0.9\linewidth]{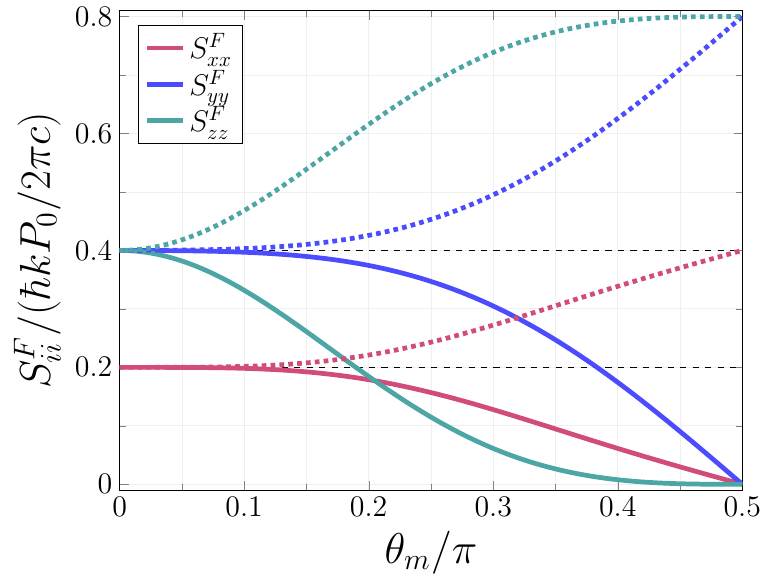}
	\caption{Backaction $S_{ii}^{F}(\omega)$ against spherical mirror spanning angle found by solving Eq.~\eqref{backaction final}. Filled lines indicate backaction for $kr_{m}=n\pi$ while dotted lines for $kr_{m}=n\pi\pm\pi/2$. Dashed lines indicate free-space levels in a standing wave, which in our model corresponds to $\theta_{m}=0$.}
	\label{fig:SWNA}
\end{figure}
\hspace{-0.12cm}where we introduced the dyadic Green's function of the system (see appendix section \ref{sec:green}), defined the free-space scattered power $P_{0}=E_{0}^{2}\alpha^{2}k^{4}c/(12\pi\epsilon_0)$ and the partial derivatives $\partial_{i}\equiv \partial/\partial x_{i}$ and $\partial_{i}'\equiv \partial/\partial x_{i}'$ with respect to the first and second argument of the Green's function respectively. As long as the aforementioned approximations apply, equation \eqref{sd mono final} represents the dominating term of backaction noise experienced by the particle in a driving laser field polarised along $\hat{\mbf{x}}$ and an arbitrary environment described by $G_{xx}$. Physically, the first and second terms in equation \eqref{sd mono final} correspond to noise in the gradient force and the scattering force respectively, while the remaining terms represent the force correlations. For the case of a standing wave in equation \eqref{SWz}, at an intensity maximum, where the particle is trapped, the expression in equation \eqref{sd mono final} can be further approximated as,
\begin{equation}
	S^{F}_{ii}(\omega)\approx\frac{3\hbar P_0}{k^2 c}\partial_i\partial_i'\im{G_{xx}(\mbf{r}_{0},\mbf{r}_{0}',\wo)}_{\mbf{r}_{0}=\mbf{r}_{0}'}
\label{fcsw mirror}
\end{equation}
leaving only the noise in the gradient force. In general, the gradient of the driving field in equation \eqref{sd mono final} gives rise to additional terms in the force spectral density. However, in the case of a particle trapped about an intensity maximum of a standing-wave, with the assumption that $|\mbf{r}_0|\ll\lambda$, the noise in the scattering force and the force correlations do not contribute up to second order in particle position \cite{seberson_2020}. In the case of a particle trapped at the focus of a single focused beam in free-space, we recover the expressions for optimal position detection found in \cite{tebben_2019} including the terms which contain the factor $A$ corresponding to the gradient of the focused field. The remaining term in equation \eqref{fcsw mirror} corresponds to the spectral density of the field gradient fluctuations. Within our approximations, our expression agrees with the expression for the decoherence rate due to laser recoil \cite{sinha_2020} for a model of a particle trapped in front of a flat plane, in the framework of macroscopic QED \cite{buhmann_2007,martinez_2022}. In that context, the dominant contribution to the noise arises from the driven-Casimir--Polder interaction term in the Hamiltonian of the trapped particle dynamics. This term is a result of interference of the classical driving field with medium-assisted vacuum fields. In addition, we note that for a standing wave under our approximations, the orientation of the beam axis does not enter the expression. Therefore, more generally the same expression holds for a standing wave rotated about the polarisation vector.

We find the effect of the mirror on the force spectral density by substituting into equation \eqref{fcsw mirror} the Green's function which characterises the spherical mirror geometry, as discussed in appendix \ref{sec:green}. The resulting total Green's function is given by equation \eqref{g decomp}, which upon substitution into equation \eqref{fcsw mirror} yields,
\begin{equation}
	S^{F}_{ii}(\omega)=\frac{\hbar k P_0}{2\pi c}\left[\frac{1}{5}(2 - \delta_{ix}\delta_{xi})-2\cos(2kr_m)\int_{(\text{a})}\d S_{ii}\right]
\label{backaction final}
\end{equation}
which represents the modified backaction noise of a particle trapped in a standing-wave at the centre of a spherical mirror. In equation \eqref{backaction final} $\delta_{xi}$ and $\delta_{ix}$ represent the Kronecker delta and we introduced,
\begin{equation}
	\d S_{ii} = (\hat{r}_i)^2\rho_x(\theta,\phi) \d\Omega
	\label{integration element}
\end{equation}
with the radiation pattern of an $\hat{\mbf{x}}$ oriented dipole,
\begin{equation}
    \rho_x(\theta,\phi) = \frac{3}{8\pi}[1-(\hat{\mbf{x}}\cdot\hat{\mbf{r}})^2].
\end{equation}
where the radial unit vector is given by $\hat{\mbf{r}}=(\sin\theta\cos\phi,\sin\theta\sin\phi,\cos\theta)$. The first term in equation \eqref{backaction final} represents the free-space backaction in a standing-wave, while the second represents the contribution of the mirror. The integration in the second term runs over the mirror surface as depicted in figure \ref{fig:measurement-zones}. For a full hemispherical mirror, the result in equation \eqref{backaction final} simplifies to,
\begin{equation}
	S^{F}_{ii}(\omega)=\frac{2}{5}\frac{\hbar k P_0}{2\pi c}\sin^2(kr_m)(2 - \delta_{ix}\delta_{xi})
	\label{backaction hemi}
\end{equation}
which shows that for $kr_m=n\pi$ with integer $n$ the dominating contribution to backaction noise vanishes. This \textit{backaction suppression condition} coincides with the condition on the mirror radius needed for trapping at the mirror centre (see appendix section \ref{sec:retro}) and is also the condition giving maximum enhancement of scattered power (see appendix section \ref{sec:emission}). This is in contrast to recent findings for the spherical mirror system \cite{weiser_2025} in which the authors assume that the suppression of scattered radiation leads to the suppression of backaction, based on previous backaction calculations in the literature for a particle scattering in free space \cite{tebben_2019}. Our result suggests that to effectively implement the protocol of \cite{weiser_2025}, which inhibits the backaction in the control region, requires tuning the relationship between the mirror radius and the wavelength to yield maximum enhancement of scattered radiation, instead of radiation suppression. In addition, our result shows that the full backaction suppression in three dimensions is enabled in a standing-wave trapping field, in which case the equation \eqref{fcsw mirror} is valid and the phase gradient of the driving field does not lead to additional contributions to backaction \cite{seberson_2020}. While the spherical geometry facilitates suppression of field gradient fluctuations, a standing wave trapping configuration is necessary to suppress residual field amplitude fluctuations. In the next section, we consider the minimum measurement imprecision achievable in a interferometric measurement of the particle's position.

In figure \ref{fig:SWNA} we also show how the backaction noise in \eqref{backaction final} varies with the mirror polar angle $\theta_{m}$, and find that the rate at which backaction decreases with $kr_{m}=n\pi$ is faster for $S_{zz}^{F}$ than for $S_{xx}^{F}$ and $S_{yy}^{F}$. This can be understood by considering the distribution of position information radiated into the far-field, which we introduce in the next section.

\section{Measurement imprecision}
\begin{figure}
    \centering
    \includegraphics[width=0.6\linewidth]{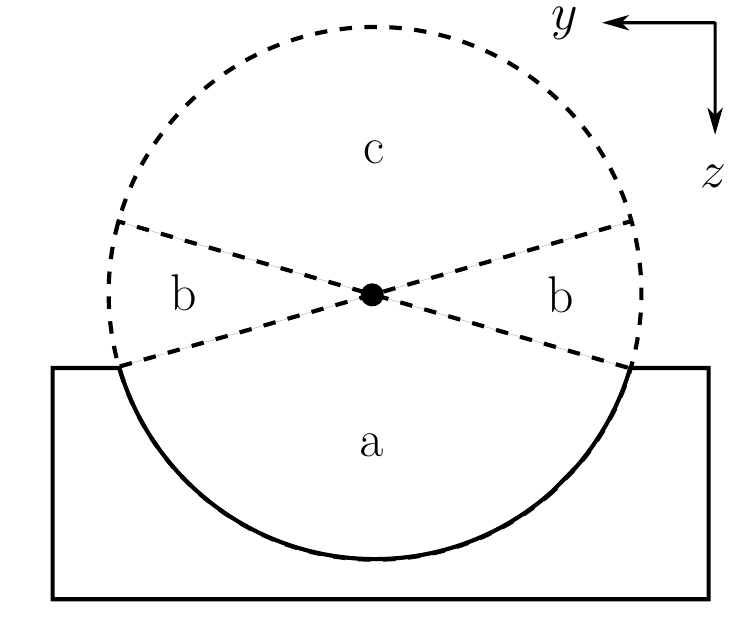}
    \caption{Diagram denoting different measurement domains. For a spherical mirror that is perfectly reflecting, no field gets transmitted and dipole emission is not accessible for measurement within domain (a). Because dipole scatter is retroreflected by the mirror surface, the mirror only affects the detected emission in domain (c), while domain (b) corresponds to free-space dipole emission. In domains (b) and (c), the dipole emission fields are given by equations \eqref{free-space dipole emission} and \eqref{dp SW 2} respectively.  }
    \label{fig:measurement-zones}
\end{figure}

\begin{figure*}
  \includegraphics{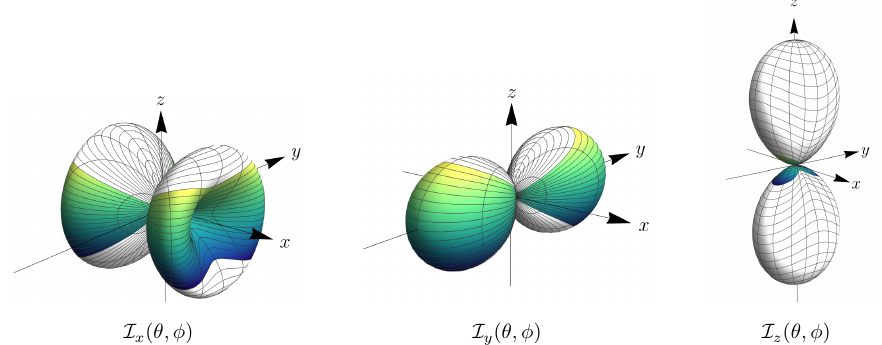}
  \caption{Angular spread of position information, here defined as the integrand of equation \eqref{Simp final} normalised to the free-space imprecision in a standing wave $S_{ii}^\text{imp}(\theta_m=0)$.  The mirror is perfectly reflecting, has its axis of symmetry along $\nz$ as in figure \ref{fig:swtrap} and was chosen to have $\theta_{m}=\pi/3$. The white region corresponds to both domain (a) and domain (c) in figure \ref{fig:measurement-zones}. The information in the white region of the top half-space (domain (a)) is never accessible because this portion of the solid angle is covered by the mirror, while information in domain (c) is rendered inacessible by our suppression scheme. At $kr_{m}=n\pi$ the position information is inaccessible in the white region of both half-spaces (a) and (b). The colored region corresponds to free-space information in a standing wave.}
  \label{fig:info patterns}
\end{figure*}
In recent studies involving the spherical mirror geometry \cite{selfhomo_2021,weiser_2025}, the achievable measurement imprecision is analysed in the context of a self-homodyne measurement; that is, a measurement in which the reflection of the particle emission acts as the reference field. This approach does not reveal the imprecision which is generally achievable in the spherical mirror geometry, when a strong, mode-matched reference field is available \cite{tebben_2019}. In this study, we formally quantify minimum measurement imprecision in the far-field, shot-noise-limited detection of particle position using the recently developed formalism of Fisher information (FI) flow \cite{fisher_2024}, which places a lower bound on the variance in the measured quantity. Specifically, the formalism introduces a local quantity of Fisher information flow, akin to the Poynting vector which describes the flow of electromagnetic energy. In the time average, the reciprocal of this quantity then places a bound on the average rate at which the variance of a variable can be reduced in a given bandwidth of shot-noise-limited measurement. The treatment is equivalent to treating the problem of an ideal interference experiment with a strong, mode-matching reference field \cite{fisher_2024,tebben_2019}. The formalism reproduces the same results but avoids introducing an ideal reference field \cite{tebben_2019}, by instead considering small perturbations to particle position via spatial derivatives. In a 1 Hz bandwidth we have,
\begin{equation}
    \text{Var}(x_{i})\geq\left(2\pi\int\mbf{S}^{\text{FI}}_{ii}\cdot\nr \d A\right)^{-1}2\pi[\text{Hz}]
\label{variance}
\end{equation}
where,
\begin{equation}
    \mbf{S}^{\text{FI}}_{ii}=\frac{2}{\hbar \omega_0}\re{\partial_{i}\mbf{E}_s\times\partial_{i}\mbf{H}_s\cj}
\label{FI flux}
\end{equation}
defines the time-averaged FI flux for the far-fields $\mbf{E}_s$ and $\mbf{H}_s$ with $\partial_{i}\equiv \partial/\partial x_{i}$, and the quantity in parentheses in equation \eqref{variance} corresponds to the flat spectral density of minimum imprecision noise,
\begin{equation}
    S^\text{imp}_{ii} = \left(2\pi\int\mbf{S}^{\text{FI}}_{ii}\cdot\nr \d A\right)^{-1}
    \label{eq:S_imp_general}
\end{equation}
in the measurement of $x_{i}$. To obtain an optimal estimate of the particle's position for general mirror spanning angle, we need to find $\mbf{S}^{\text{FI}}_{ii}$ in each angular domain at some large distance $r_d$ over the angular element $\d A = r_{d}^{2}\d\Omega$, as depicted in figure \ref{fig:measurement-zones}. In domain (b), the dipole emission is unaffected by the mirror and hence the electric field is equal to the free-space dipole emission field driven by the field in equation \eqref{SWz},
\begin{equation}
    \mbf{E}_s^{(b)}(\mbf{r}_{d},\mbf{r}_{0}) = \mbf{E}_{s}(\mbf{r}_{d})\cos(Akz_0)e^{-ik\nr\cdot\ro}
\label{free-space dipole emission}
\end{equation}
where we expanded $r'=|\mbf{r}_d-\mbf{r}_0|\approx r_d -\hat{\mbf{r}}\cdot\mbf{r}_0$ and where $\hat{\mbf{r}}=(\sin\theta\cos\phi,\sin\theta\sin\phi,\cos\theta)$. The fields in equation \eqref{free-space dipole emission} $\mbf{E}_s^{(b)}(\mbf{r}_d,\mbf{r}_0)$ and $\mbf{E}_s(\mbf{r}_d)$ represent the complex field amplitudes in domain (b) from a dipole oriented along $\hat{\mbf{x}}$ and located at position $\mbf{r}_0$ and the origin respectively (see also equation \eqref{far-field-emission}). In domain (c), the fields are a sum of the free-space dipole emission and its reflection from the spherical mirror. For a perfectly reflecting mirror and small displacement of the dipole from the mirror centre, we find the field in domain (c) to be, \footnote{See section 1 in the supplementary material for a detailed derivation.}
\begin{equation}
    \E_{s}^{(c)}(\mbf{r}_{d})=\E_{s}(\mbf{r}_{d})e^{ikr_{m}}\cos(Akz_{0})\cos(k[r_m + \hat{\mbf{r}}\cdot\mbf{r}_0]).
\label{dp SW 2}
\end{equation}
Note that in the field in equation \eqref{dp SW 2} the free-space dipole emission and its reflection are in-phase. This is because the reflected field undergoes a change of sign upon propagating through the focus from the mirror surface to domain (c), akin to the Gouy shift of a focused Gaussian beam \cite{Feng_01}. 

Substituting these fields into equation \eqref{eq:S_imp_general} and evaluating the derivatives about the mirror centre yields the final result for a general spanning angle of the mirror,
\begin{equation}
    S^{\text{imp}}_{ii}=\left(\frac{8\pi k P_0}{\hbar c}\left[\int_{(\text{b})}\d S_{ii}+4\sin(kr_{m})^2\int_{(\text{c})} \d S_{ii}\right]\right)^{-1}
\label{Simp final}
\end{equation}
where $\d S_{ii}$ is given by equation \eqref{integration element}. In the absence of the mirror (vanishing of domain (c)), the expression in equation \eqref{Simp final} recovers the expressions for free-space from \cite{tebben_2019} with the exception of the term proportional to the factor $A$ along the beam propagation direction. This is because in a standing-wave field, the field gradient does not carry linear position information. For a full-hemisphere, the result in equation \eqref{Simp final} reduces to,
\begin{equation}
    S^{\text{imp}}_{ii}=\frac{5}{8}\frac{\hbar c}{\pi k}\frac{1}{P_0}\sin(kr_m)^{-2}(1 + \delta_{ix}\delta_{xi}).
\label{simp hemi}
\end{equation}
Equation \eqref{simp hemi} diverges at the backaction suppression condition found in previous section, suggesting that the emission from the particle available for measurement contains no first-order position information. Indeed, when $kr_m=n\pi$ the total emission field in the available solid angle,
\begin{equation}
    \E_{s}^{(c)}(\mbf{r}_{d},\mbf{r}_0)\approx 2\Es(\mbf{r}_{d})
\end{equation}
is independent of the source position to first order; the free-space dipole emission is in-phase with the emission from the image dipole \footnote{The image of a point scatterer at $\mathbf{r}_0$ near the centre of a spherical mirror appears at the reflected position $-\mathbf{r}_0$, for $|\mathbf{r}_0|\ll\lambda$.}, making it impossible to distinguish between them to first order.

Let us also remark on the distribution of available position information in the context of the angular information patterns which are employed in the literature \cite{tebben_2019,ballestero_2023}. In figure \ref{fig:info patterns} we show the distribution of radiated position information for a spherical mirror with an illustrative spanning angle of $\theta_m=\pi/3$. The shapes of the corresponding information patterns are unchanged from free-space patterns in a standing-wave, but the amount of information changes discontinuously between angular domains depicted in figure \ref{fig:info patterns}. Note that the distribution of the $\mathcal{I}_z(\theta,\phi)$ information differs from that given in \cite{tebben_2019}; in a standing wave field, the information about the position along $\hat{\mbf{z}}$ is radiated symmetrically about the $xy$ plane. For a perfectly reflecting mirror at $kr_m=n\pi$, the position information is inaccessible in domains (a) and (c). In addition, figure \ref{fig:info patterns} shows that at $\theta_m=\pi/3$ a significantly larger portion of the $\mathcal{I}_z$ solid angle is affected by the mirror compared with $\mathcal{I}_x$ and $\mathcal{I}_y$. The spherical mirror with its axis of symmetry aligned along the axis of the standing wave, is particularly well-suited for suppression of position information along the axis. This is also visible in the corresponding plot of backaction noise against $\theta_m$ in figure \ref{fig:SWNA}, which shows that the noise along $\hat{\mbf{z}}$ changes more rapidly with $\theta_m$ than along $\hat{\mbf{x}}$ and $\hat{\mbf{y}}$.

\section{Discussion}
We find that the product of the imprecision in Eq.~\eqref{simp hemi} and the backaction in Eq.~\eqref{backaction hemi} satisfies the Heisenberg limit of detection, \cite{tebben_2019,clerk_2010}
\begin{equation}
    S_{ii}^F S_{ii}^\text{imp} = \left(\frac{\hbar}{2\pi}\right)^2
\end{equation}
for any choice of mirror radius. The limit is also satisfied by the solutions in equation \eqref{Simp final} and \eqref{backaction final} for an arbitrary choice of mirror spanning angle $\theta_m$. This results suggests that by utilising a spherical mirror one can make a trade-off between measurement imprecision and backaction by choosing an appropriate mirror radius. For a perfectly reflecting hemisphere at $kr_{m}=n\pi$, the detected scatter is maxiamlly enhanced, but contains no linear position information. Correspondingly, the dominating term of backaction noise vanishes. This reinforces the fact that the flow of energy does not necessarily carry information about the environment with which it has interacted \cite{fisher_2023,tebben_2019}. 

The suppression effect could be tested experimentally by measuring the reheating rates via trapped particle trajectories, as was done in free space \cite{Jain_2016}. The necessary linear measurement of particle position could be accessed with illumination in the $xy$ plane for a mirror with a spanning angle $\theta_m<\pi/2$, or using a detection beam which does not satisfy the suppression condition for the mirror radius ($kr_m\neq n\pi$). In addition, we note that second-order position information should be accessible under the suppression condition, which is sufficient for the implementation of parametric feedback cooling \cite{Gieseler_2012} or temperature measurement.

The largest contributions to residual backaction noise are likely to arise from technical limitations in the experimental realisation. According to equation \eqref{gs-intro}, imperfect reflectivity of a metallic mirror affects the suppression result in equation \eqref{backaction final} linearly, and would limit the suppression of backaction noise to about $10^{-2}$ of its free-space magnitude. A realisation scheme will also require surface quality and long-term mirror radius stability $\delta r_m \ll \lambda/2\pi$. Considering the coefficient of linear expansion of aluminium $\alpha_\text{Al}=2.3\times 10^{-5}\text{K}^{-1}$ and a mirror with radius of about 1 mm, sufficient radius stability can be achieved with thermal stability $\delta T < 1\text{K}$. A recent experimental proposal utilising a spherical mirror for similar application suggests that such temperature stabilisation and the quality of the fabricated mirror surface should be achievable \cite{panopticon_2020}.

\section{Conclusion}
In summary, we have found that a suitable structured environment can yield significant reduction of mechanical noise due to backaction. We have studied a spherical mirror as a particular case, and found that it can be used to inhibit the backaction force noise in three dimensions. The suppression scheme also requires the particle to be trapped in a standing wave trapping field. We have found that under a suitable condition on the mirror radius, the mirror significantly attenuates the attainable acquisition rate of linear position information, at the same time suppressing the largest backaction noise term. In the case that the mechanical noise experienced by the particle is dominated by backaction and the detection is shot-noise limited, we have shown that the imprecision and backaction noise satisfy the Heisenberg limit of detection for any choice of mirror radius $r_m$ and spanning half-angle $\theta_m\leq\pi/2$. Experimental considerations show that the suppression scheme should be achievable in a realistic setting; the amount by which recoil noise can be suppressed using our method will either be limited by experimental factors (such as mirror reflectivity, thermal stability, or surface quality) or smaller residual backaction noise terms unaccounted for in our calculation. 

Counter-intuitively, our results show that the suppression condition is also the condition for which the scattered power, usually associated with recoil noise, is maximally enhanced by the spherical mirror. This result suggests that a scheme using a spherical mirror to constrain the backaction noise such as the one analysed in \cite{weiser_2025}, should aim to maximise scattered radiation instead of suppressing it. Since the scattered power of a particle radiating at the centre of a spherical mirror has a sinusoidal dependence on the mirror radius, this change to the protocol of \cite{weiser_2025} can be effected with a small adjustment of the mirror radius by $\lambda/4$.

Our result can be understood in the context of the position information content in the outgoing particle scatter; at the suppression condition, the image dipole appears identical to the real particle to first order in particle position about the mirror centre, preventing localisation in a position tracking experiment. For all experimental conditions considered, we have shown that this measurement imprecision and the corresponding backaction satisfy the Heisenberg limit of detection. The method presented here shows how an appropriate reflective surface geometry can be used to mitigate backaction, and may enable development of more sophisticated schemes for particles of different sizes and shapes in future experiments. In addition, a wider range of geometries may be suitable, if one wishes to suppress the backaction noise only in one or two directions.

The main result shows a strong dependence of the imprecision-backaction characteristics on the surrounding mirror geometry. The spherical mirror geometry is uniquely suited for point-like particles as the mirror reflection can perfectly match dipole emission away from the mirror. Other geometries may present similar useful properties in future investigations, such as a plane mirror-lens system. Other practical arrangements suitable for the scattering properties of particles with different and more complex morphologies may be found by adapting existing analysis tools, used for tailoring the optical properties of nanoscale emitters \cite{Mignuzzi_2019,Barnes_2020}.  We once again note that although the spherical mirror geometry plays a necessary role for the suppression effect, it is not sufficient for full suppression in three dimensions; a standing wave trapping potential is also required to remove the contribution of the local phase gradient, which would be present in a single-beam configuration \cite{tebben_2019}.

\begin{acknowledgments}
  R.G. was supported by the UK Engineering and Physical Sciences Research Council (EPSRC) through Standard Research Studentship (DTP) EP/T517987/1.
  No new data were generated or analysed during this study.
\end{acknowledgments}
\appendix
\section{System's dyadic Green's function}\label{sec:green}
% \input{green}
Green's functions are solutions to inhomogeneous differential equations with the inhomogeneity given by a delta function. In homogeneous space, a point dipole gives a singular inhomogeneity in the Helmholtz equation. Fields radiated by a point-like electric dipole are determined by the \textit{dyadic Green's function} $\G$; a tensor of rank two which solves the equation, \cite{novotny_2012principles}
\begin{equation}
  \nabla\times\nabla\times\G(\mbf{r},\mbf{r}_0,\omega) - k^{2}\G(\mbf{r},\mbf{r}_0,\omega) = \II\delta(\mbf{r}-\mbf{r}_0)
\end{equation}
with wavenumber $k=\omega/c$, unit dyad $\II$, source position $\mbf{r}_0$ and observation point $\mbf{r}$. In free space, the solution for the Green's function is, \cite{novotny_2012principles}
\begin{equation}
  \Go(\mbf{r},\mbf{r}_0,\omega) = \left[\II + \frac{1}{k^{2}}\nabla\nabla\right]\frac{\exp(ikr')}{4\pi r'}
\label{gfree}
\end{equation}
where we wrote $r'=|\mbf{r}'|=|\mbf{r}-\mbf{r}_0|$ and a direct product of two vectors corresponds to a dyadic product. In the subsequent calculations, we will make particular use of the part of $\Go$ which scales as $(r')^{-1}$,
\begin{equation}
  \G_\text{ff}(\mbf{r},\mbf{r}_0,\omega) = \frac{\exp(ikr')}{4\pi r'}\left[\II - \frac{\mbf{r}'\mbf{r}'}{(r')^{2}}\right]
\label{g_far_field}
\end{equation}
which is gives rise to the far-field free-space dipole emission given by,
\begin{equation}
  \mbf{E}_s(\mbf{r},\mbf{r}_0) = \frac{k^2}{\epsilon_0}\dyad{G}_\text{ff}(\mbf{r},\mbf{r}_0,\omega)\cdot\mbf{p}
  \label{far-field-emission}
\end{equation}
where $\mbf{p}$ is the electric dipole moment. 
We can also use the system's dyadic Green's function element corresponding to the dipole orientation to evaluate the correlation functions of the field fluctuations in the framework of stochastic electrodynamics \cite{sideband_2022,novotny_2012principles}. At thermal equilibrium at temperature $T$ this fluctuation-dissipation theorem for the $\hat{\mbf{x}}$ component of the field fluctuations reads, 
\begin{equation}
	\avg{\delta\ft{E}\cj(\mbf{r},\omega)\delta \ft{E}(\mbf{r}',\omega')} =f(\mbf{r},\mbf{r}';\omega,T)\delta(\omega-\omega'),
	\label{fdt}
\end{equation}
where,
\begin{equation}
	f(\mbf{r},\mbf{r}';\omega,T)=\frac{\hbar\omega^{2}}{2\pi c^{2}\epsilon_0}\coth\left(\frac{\hbar\omega}{2k_B T}\right)\im{G_{xx}(\mbf{r},\mbf{r}',\omega)}.
\end{equation}
In the main text, we use equation \eqref{fdt} to model the spectral density of force fluctuations due to backaction.

In the presence of reflective boundaries, the total Green's function can be decomposed as,
\begin{equation}
    \G(\mbf{r},\mbf{r}_0,\omega)=\G_0(\mbf{r},\mbf{r}_0,\omega)+\Gs(\mbf{r},\mbf{r}_0,\omega)
\label{g decomp}
\end{equation}
where the first term corresponds to free-space emission, and the second term corresponds to the \textit{reflected} field and depends on the boundary. For a dipole emitter located at or near the centre of a large spherical mirror (with radius $r_m\gg\lambda$), only the far-fields contribute to the reflected fields near the source position. Modelling the reflected fields using Fresnel coefficients and the angular spectrum representation for focused far-fields \cite{novotny_2012principles, mandel_wolf_1995}, at an observation point $\mbf{s}$ at or near the centre of the mirror we find the expression, 
\begin{equation}
  \Gs(\mbf{s},\mbf{r}_{0},\omega) = -\rho\frac{ikr_{m}e^{ikr_{m}}}{2\pi}\int_{(\text{a})}\dyad{G}_{\text{ff}}(\mbf{r},\mbf{r}_{0},\omega)e^{-ik\nr\cdot\mbf{s}}d\Omega
    \label{gs-intro}
\end{equation}
where $k=\omega/c$, $\rho$ denotes the Fresnel reflection coefficient of the mirror at normal incidence, $\mbf{r}$ denotes a vector from the origin to a point on the mirror surface, and the integration of the solid angle element $\d\Omega$ runs over the solid angle of the mirror (denoted as domain (a) in figure~\ref{fig:measurement-zones})

\section{Dipole emission near boundaries}\label{sec:emission}
The scattering rate of a point dipole emitter depends on its environment. For a point dipole emitter, the average scattered power can be expressed as, \cite{Barnes_2020,novotny_2012principles}
\begin{equation}
  \avg{P} = \frac{6\pi P_0}{k}\mbf{n}_p\cdot \im{\G(\mbf{r}_0,\mbf{r}_0,\omega_0)}\cdot\mbf{n}_p
\label{scatter_mod}
\end{equation}
where the dyadic Green's function is evaluated at the location of the dipole in both arguments, $P_0$ denotes he free-space scattered power, and the vector $\mbf{n}_p$ denotes the dipole orientation.
\begin{figure}
	\centering
	\includegraphics[width=0.5\textwidth]{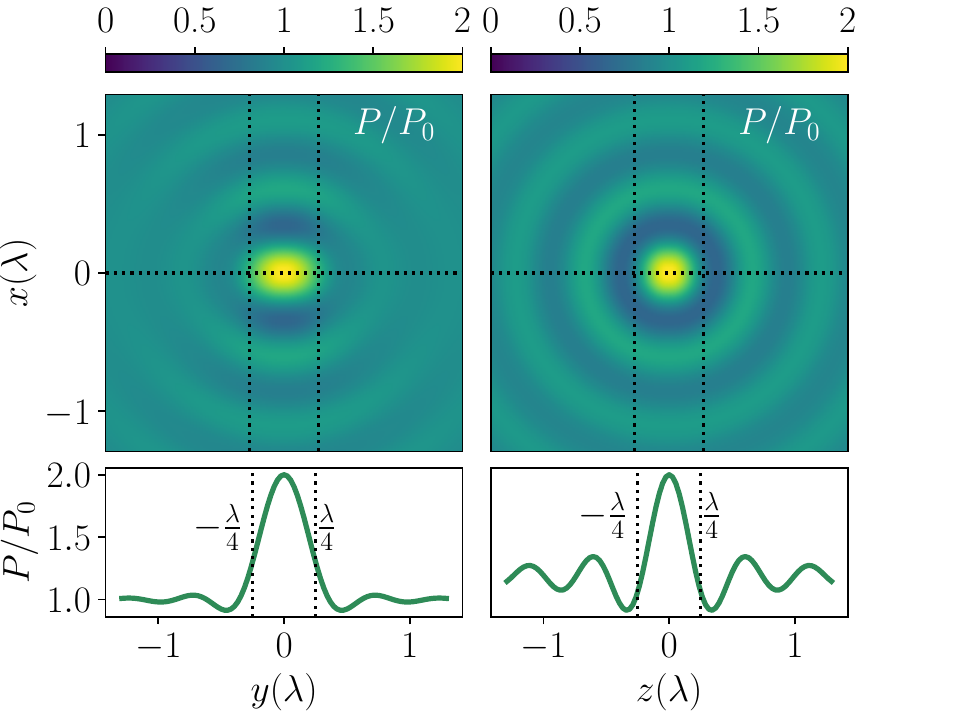}
	\caption{Modified scattered power, expressed as a ratio with the free-space scattering rate $P/P_0$, evaluated at the mirror radius condition $kr_m=n\pi$. The modified scattered power for a displaced source is found by numerically solving the scattering Green's function in equation \eqref{gs-intro}.}
	\label{fig:enhancement}
\end{figure}
By substituting equation \eqref{g decomp} into \eqref{scatter_mod} and using equations \eqref{gfree} and \eqref{gs-intro}, we find that at the centre of a perfectly reflecting ($\rho=-1$) hemispherical mirror the scattered power is given by,
 \begin{equation}
    \avg{P} = 2\cos^{2}(kr_m) P_0.
\label{mod_power_origin}
\end{equation}
In agreement with the QED result found for a spontaneous emission from an atom at the centre of a hemispherical mirror \cite{Hetet_2010}; the condition $kr_m=n\pi$ corresponds to a two-fold enhancement of the total scattered power. In figure \ref{fig:enhancement} we also show that the scattered power varies as the location of the dipole is varied about the centre of the mirror.

\section{Retro-reflection from a spherical mirror}\label{sec:retro}
In this section we find the necessary conditions for retro-reflection of a laser beam from the surface of a spherical mirror, and the corresponding mirror radius which allows trapping at the mirror centre. The beam is retro-reflected when its radius of curvature matches that of the mirror. Modelling the laser as a paraxial Gaussian beam with radius of curvature $R(z)$, this condition is expressed as, \cite{davis_2014}
\begin{equation}
r_{m}=R(z-z_{f}) = z - z_{f} + \frac{z_{R}^{2}}{z-z_{f}}
\label{curvature condition}
\end{equation}
where $z_f$ is the axial position of the focus along the mirror axis. The condition in equation \eqref{curvature condition} has two solutions for the focus position,
\begin{equation}
	z_{f}^{\pm}=\frac{r_{m}}{2}\left(1\pm\sqrt{1-4\frac{z_{R}^{2}}{r_{m}^{2}}}\right).
\label{sw mirror focus solutions}
\end{equation}
Since we want the laser intensity at the trapping position to be as large as possible, we choose the negative sign solution. Therefore, we can achieve maximum modulation when the beam is focused a distance $z_{R}^{2}/r_{m}$ (for $z_R\ll r_m$) from the mirror centre towards its surface. Note that since $z_{R}\approx\lambda/\pi(\text{NA})^{2}$ for a beam of numerical aperture NA, the tighter the focus of the beam, the closer the solution $z_{f}^{-}$ is to the centre of the mirror.

To find a condition on the mirror radius which yields an intensity maximum at the centre of the mirror, we consider the round-trip phase about the mirror centre. For a plane wave reflecting from a perfect plane mirror a distance $r_{m}$ away this round-trip phase would simply be equal to $2kr_{m}+\pi$. For a Gaussian beam, there is an additional contribution arising from the Gouy phase shift $\phi_g(z) = \arctan(z/z_R)$ \cite{Feng_01}. With the focus at $z_{f}^{-}$ we find that,
\begin{equation}
	\Delta\phi = 2kr_{m} + \pi +\underbrace{2\left[\phi_{g}(z_{f})+\phi_{g}(r_{m}-z_{f})\right]}_{\pi} = 2kr_{m}.
\end{equation}
That is, the contribution of the Gouy phase shift to the round-trip phase is equal to exactly $\pi$ when the beam is retroreflected, such that the total phase is fixed only by the round-trip distance. To ensure that the incident beam is in-phase with its reflection at the centre, the mirror radius has to be restricted to $kr_m = n\pi$, for integer $n$. This condition corresponds to the maximum enhancement of scattered power from a trapped particle. As we discussed in the main text, this condition is exactly the condition needed to achieve suppression of backaction noise acting on a trapped particle. This coincidental feature ensures that the recoil noise suppression condition is matched when the particle is brought to the centre of the spherical mirror in a standing-wave, formed by reflection from the mirror.

\bibliography{main}
\end{document}